\documentstyle[aps,multicol,amsmath,graphics]{revtex}  

\newcommand{\erf}{\mbox{erf}}
   
\begin{document} 
\title{On time's arrow in Ehrenfest models with reversible
    deterministic dynamics}   
 \author{R. Metzler and W. Kinzel}
\address{Institut f\"{u}r Theoretische Physik, 
Universtit\"{a}t W\"{u}rzburg, Am Hubland, D-97074 W\"{u}rzburg, Germany}
\author{I. Kanter}
\address{Minerva Center and Department of Physics, Bar Ilan University, 52900 
Ramat Gan, Israel} 
\maketitle
 
\begin{abstract}  
We introduce a deterministic, time-reversible version 
of the Ehrenfest urn model. The distribution of first-passage
times from equilibrium to non-equilibrium states and vice
versa is calculated. We find that average times for transition 
to non-equilibrium always scale exponentially with the system size,
whereas the time scale for relaxation to equilibrium depends
on microscopic dynamics. To illustrate this, we also look at
deterministic and stochastic versions of the Ehrenfest model 
with a distribution of microscopic relaxation times.   
\end{abstract}

\begin{multicols}{2}

\section{Introduction} 
Physical laws of motion are time reversible. As a consequence, from a
movie of a few interacting particles one cannot distinguish whether the
movie is running forward or backward. This seems to be different when we
observe a very large number of particles. Macroscopic properties like
the density of particles show a direction of time. If the milk in the
coffee shrinks to a single drop we know that the movie is running
backward.
 
More than a hundred years ago Ludwig Boltzmann gave convincing arguments
for the irreversible macroscopic behaviour of particles moving with
reversible microscopic laws. The fraction of initial states which leads
to reversible macroscopic behaviour is so extremely small that we will
never be able to observe it. With probability one, we observe a time's
arrow in macroscopic properties.
 
Hence, through the selection of the initial state, the deterministic
motion receives probabilistic elements. In fact describing a
large system of interacting particles by either a stochastic motion
(Boltzmann transport equations) or ensemble theory (statistical
mechanics) turned out to be very successful.
 
However, the success of Boltzmann's ideas did not suppress the
discussion about the foundations of irreversible properties, even
hundred years later 
\cite{Lebowitz:Macroscopic,Lebowitz:Entropy,Lebowitz:More,Stauffer:Q2R,Ambegaokar:Entropy}. 
For example, there was a round table
discussion on this subject on the STATPHYS 98 conference 
\cite{Lebowitz:Microscopic,Prigogine:Laws},
where Prigogine introduced novel microscopic laws which are irreversible
with time. One reason for this ongoing discussion is the absence of
rigorous mathematical proofs of irreversible properties in the
thermodynamical limit. Furthermore, solvable models where irreversible
macroscopic properties can be well defined and investigated almost do
not exist, to our knowledge. Either one studies stochastic models like
the famous one of Paul and Tanja Ehrenfest \cite{Ehrenfest:Model}, 
or disorder averages
as for the Kac ring \cite{Kac:Probability}, 
or ensembles of chaotic noninteracting
particles as in the Lorentz gas \cite{Lebowitz:Lorentz,Lebowitz:Lorentz2}. 
However, a single trajectory of
interacting particles is difficult to investigate in this context. 

Numerical simulations of the equations of motions do not help much in
this context, either. The algorithms can be formulated in a time
reversible way and the decay of Boltzmann's H--function has been
demonstrated \cite{Levesque:Dynamics}. However, the state space is too
large to study any details, for instance to estimate the fraction of
untypical inital states or to find the time scale of the Poincar\'e
return time.

One should keep in mind that we want to understand the property of a
{\it single} trajectory for a {\it short} time. Hence, ensemble averages
do not give a basic explanation of irreversible properties, since they
contain an average over infinitely many trajectories. Ergodic theory
does not help either, since it needs time averages over infinitely large
times (at least up to the Poincar\'e return time) whereas real physical
systems or computer simulations run over time scales which are extemely
short compared to the return time.
 
In this paper we introduce a model with deterministic time reversible
dynamics which can be analysed in detail. It is a deterministic
extension of the Ehrenfest model discussed in Section
\ref{StochEhr}. The dynamics is ergodic, each state is
visited once during a cycle. The Poincar\'e return time is known exactly
and the Boltzmann and Gibbs entropies can be defined precisely. From
numerical calculations we obtain the distribution of first passage times
the system needs to go from a non equilibrium state to a state with
largest Boltzmann entropy as well as in the reverse direction.
The scaling of average first passage times with system size is
calculated. Finally we compare the results of the deterministic model
with the corresponding ones of the stochastic version of the Ehrenfest
model.
\section{Deterministic Ehrenfest model: shift register generator} 
The microscopic state $x$ of our model has $N$ binary variables $x_i \in
\{0,1\}; i = 1,..., N$,
\begin{equation}\label{eins} 
x(t) = (x_1, x_2, ... , x_N) \ \ . 
\end{equation}  
The time $t$ is discrete. At each time step a new bit $x_0$ is generated
and replaces $x_1$, whereas the old bits $x_1,\dots, x_{N-1}$ are moved
one step to the right:
 
\begin{eqnarray}\label{zwei} 
& x_1 & \rightarrow  x_2 \nonumber\\ 
& \vdots  & \nonumber \\ 
& x_{N-1} & \rightarrow x_N \nonumber \\ 
& x_0 & \rightarrow  x_1\ \ . 
\end{eqnarray} 
The last bit $x_N$ is deleted. The new bit $x_0$ is constructed using
the theory of primitive polynomials modulo two which is also used to
generate pseudo random numbers 
\cite{Knuth:Art2,NumRec,Schroeder:Zahlentheorie,Ziff:Four-tap}. 
If $x_0$ is the sum modulo two of a few bits
$x_i$ at certain positions $i$, then any initial state $x(0)$ runs
through all possible states $x$ except the state zero $(0, 0, ..., 0)$.
For instance for $N= 97$ one finds from the table of Ref. \cite{NumRec}:
\begin{equation}\label{drei} 
x_0 = (x_{97} + x_6) \bmod \; 2\; . 
\end{equation} 
In our investigation we use such sequences
with maximal cycle length, taking the generators from \cite{NumRec}
for $N<100$ and from \cite{Ziff:Four-tap} for higher $N$.
 
For the macroscopic property $M(x)$, which we will investigate, we take
the number of $x_i = 1$,
\begin{equation}\label{vier} 
M(x) = \sum\limits^{N}_{i=0} x_i \ \ . 
\end{equation} 
Note that at each time step $M$ changes by $\pm 1$ at most. The original
Ehrenfest model considered $N$ balls in two urns. $x_i = 0(1)$ means
that ball $i$ is in the left (right) urn. $M$ is the number of balls in
the right urn. In the Ehrenfest model the balls were chosen randomly.
Here we define a deterministic rule to move the particles between the
two urns.
 
The equations of motion (\ref{zwei}) and (\ref{drei}) are time
reversible; for instance from (\ref{drei}) one finds
\begin{equation}\label{fuenf} 
x_{97} = (x_0 + x_6) \bmod\;  2\; . 
\end{equation} 
Hence, from the first and seventh bit of state $x(t+1)$ one calculates
the 97$^{th}$ bit of $x(t)$, and a shift to the left gives the rest of
$x(t)$.
 
We have now defined a model where time and states are discrete. It has
the following properties:
\begin{enumerate} 
\item The equations of motion are deterministic and time reversible. 
\item Each initial state returns to itself after $2^N - 1$ time steps;
  i.e. the Poincar\'e return time is $T_R = 2^N - 1$.
\item The Boltzmann entropy $S_B$ is given by the number of microstates
  $x$ which have the macroscopic property $M$. Since the distribution
of macrostates is the binomial distribution (with the exception of 
the zero state),
\begin{equation}
p(M(x)) = \binom{N}{M(x)}/(2^N-1),
\end{equation}
one finds for the entropy in dimensionless units:
\begin{equation}\label{sechs} 
S_B (x) = \ln \binom{N}{M(x)} \ \ . 
\end{equation} 
\item Since each state $x$ is visited once during each cycle, it has an
  identical  statistical weight in a Gibbs ensemble. Hence, the Gibbs
  entropy $S_G$ is given by
\begin{equation}\label{sieben} 
S_G = \ln (2^N - 1) \simeq N\ln 2 \ \ . 
\end{equation} 
\item The system is ergodic. When the system is observed over the return
  time $T_R$, the time average $\overline{M} $ agrees with ensemble 
average $\langle M \rangle$:
\begin{equation}\label{acht} 
\overline{M} = \frac{1}{T_{R}} \; \sum\limits^{T_{R}}_{t>0} \;\; 
  M[x(t)] = \langle M \rangle = \frac{N}{2} \ \ . 
\end{equation} 
\item 
The most probable value of $M$ agrees with time and ensemble average, 
\begin{equation}\label{neun} 
M_{mp} = \overline{M} = \langle M \rangle = \frac{N}{2} \ \ . 
\end{equation} 
There are  
\begin{equation}\label{zehn} 
{N \choose N/2} \simeq \sqrt{\frac{2}{\pi}} \; \frac{2^N}{\sqrt{N}}  
\end{equation} 
states $x$ whith $M(x) = M_{mp}$. We call these states {\it equilibrium
  states}.
\end{enumerate} 
During a cycle, the average time interval between two 
consecutive equilibrium states increases with $\sqrt{N}$ according  
to Eq. (\ref{zehn}).
\section{Macroscopic dynamics} 
\begin{figure}
 \begin{center}
   \resizebox{0.95 \columnwidth}{!}{\includegraphics{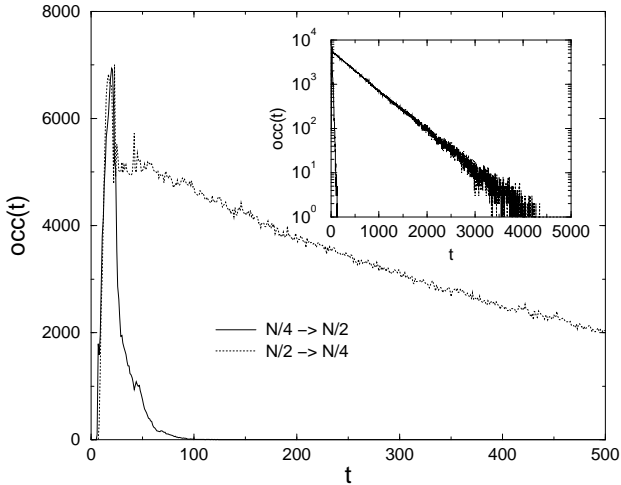}}
 \end{center}
 \caption{Distribution of first-passage times to equilibrium
($N/2$, solid line) and non-equilibrium ($N/4$, dotted line) 
in the 24-bit shift register. All states are counted; 
therefore the integral 
over the solid curve is $\tbinom{24}{6}$ while that over the
dotted curve is $\tbinom{24}{12}$. The inset shows the same
curves on a semi-logarithmic scale.}
 \label{sh-fpt}
\end{figure}

We investigate the two following questions: 
\begin{enumerate} 
\item When the system starts from an initial state $x$ which is far 
away from equilibrium, for instance with $M(x) =  
N/4$, how long does it take to reach an equilibrium 
(= most probable) state?  
\item When the system starts from an equilibrium state, 
$M(x) = N/2$, how long does it take to reach a non-equilibrium  
state with $M= N/4$? 
\end{enumerate} 
We call these two time intervals first passage time $T_{eq}$ and
$T_{neq}$, respectively. Both of these times depend on the special
choice of the initial state, therefore we obtain a distribution of first
passage times. Fig. \ref{sh-fpt} shows the result for $N=24$
which is obtained from
simulations of all $2^{24} - 1$ states.  The time $T_{eq}$ to
reach an equlibrium state has a sharp peak around $T_{eq}\simeq N$. In
contrast, the time to leave equilibrium has a broad distribution
extending from the minimal possible time $T_{neq}=N/4$ to
$T_{neq}=5079$. Since the system is time reversible,
 the distribution of equilibrium times should be part of  
the distribution of non-equilibrium times $T_{neq}$. In fact, 
we observe the corresponding peak
in Fig. \ref{sh-fpt}. But surprisingly there are more 
equilibrium states which
reach non equilibrium ones for the first time 
in short than in long times. The
distribution of $T_{neq}$ has its maximum for values of the order of $N$,
while the tail can be fitted to an exponential distribution 
\begin{equation}
P(t) \propto e^{-t/\tau},
\end{equation}
where $\tau$ is of the order of the average non-equilibrium time
$\langle T_{neq}\rangle$, averaged over all initial equilibrium states.

\begin{figure}
 \begin{center}
   \resizebox{0.95 \columnwidth}{!}{\includegraphics{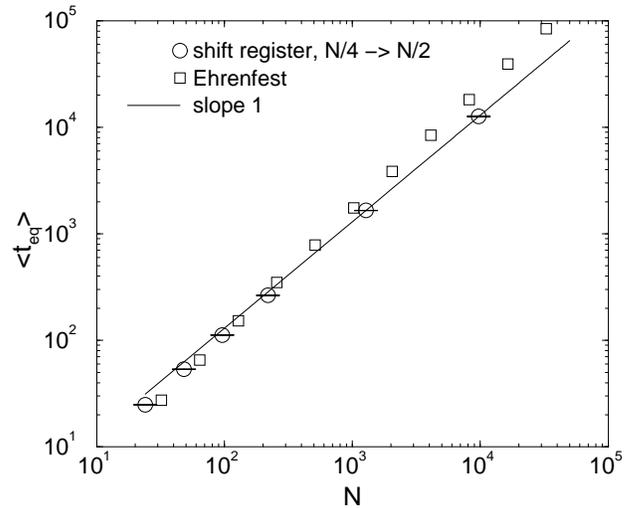}} 
 \end{center}
 \caption{Mean first-passage times from $M=N/4$ to equilibrium 
   for different sizes $N$ of the shift register and Ehrenfest
model.}
 \label{sh-rel}
\end{figure}

\begin{figure}
 \begin{center}
   \resizebox{0.95 \columnwidth}{!}{\includegraphics{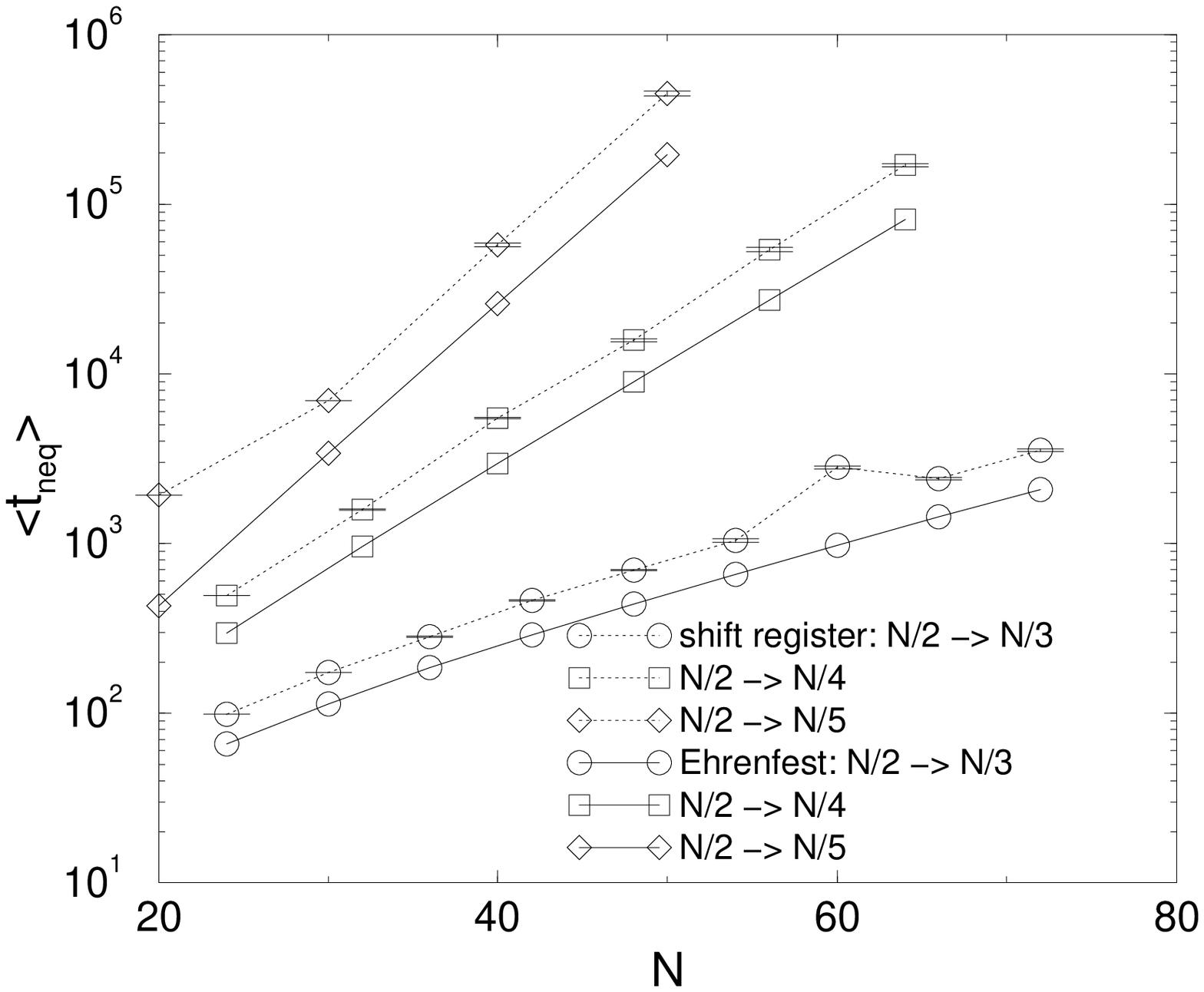}} 
 \end{center}
 \caption{Mean first-passage times from equilibrium to 
$M=N/3$, $N/4$ and $N/5$, for different $N$ of
the shift register and Ehrenfest model. Two-tap 
shift generators were used for $N=20$ and $N=60$, while
the other points are from four-tap generators.}
\label{sh-inv}
\end{figure}

\begin{figure}
 \begin{center}
   \resizebox{0.95 \columnwidth}{!}{\includegraphics{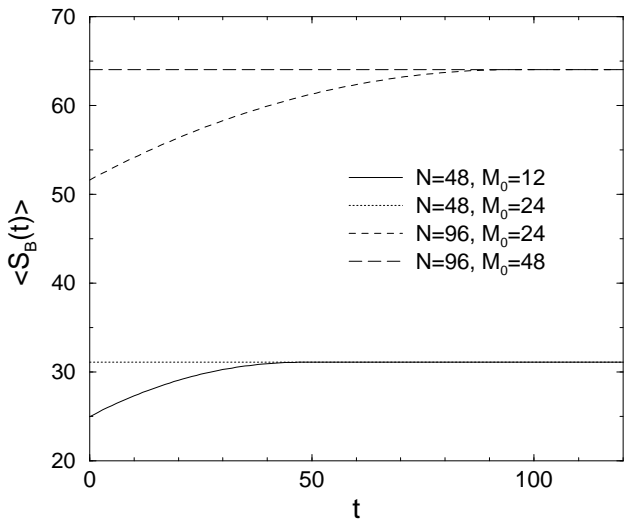}} 
 \end{center}
 \caption{Time development of the Boltzmann entropy $S_B$ of
shift registers with $N=48$ and $N=96$. Simulations averaged
over $10^5$ initial conditions with $M_0=N/4$ and $M_0=N/2$.
The entropy relaxes to its maximum value in $N$ time steps.}
 \label{sh-ent}
\end{figure}

In Figs. \ref{sh-rel} and \ref{sh-inv} it is shown how the 
average value of these times scales with
system size $N$. For small systems $N<32$ we average over all possible
initial states whereas for large systems we average over several
hundred randomly selected initial states. We obtain a good fit with
\begin{eqnarray}\label{elf} 
\langle T_{eq}\rangle & = & 1.30 N -12.3 \nonumber \\ 
\langle T_{neq}\rangle & = & 15.0 \exp(0.146 N) \ \ . 
\end{eqnarray} 
Hence, the time to reach equilibrium is short, it scales with system
size $N$, whereas the time to reach a state far from equilibrium is very
long and increases exponentially with $N$. A lower bound on
$\langle T_{neq}\rangle $ 
can be found for any system whose distribution of macrostates is
the binomial distribution: take a sequence of 
$\tbinom{N}{N/2} / \tbinom{N}{N/4}$ equilibrium states followed by one
nonequilibrium state with $M = N/4$, repeated as often as necessary. 
Incidentally, this lower bound differs only by a factor of $\sqrt{2/(\pi N)}$
from the average return time of the non-equilibrium state.
Choosing an initial occupation $M_0 = mN$, the lower bound 
can be approximated for large $N$ using Stirling's Formula:
\begin{equation}
\langle T_{neq} \rangle \ge \sqrt{2 m(1-m)} (2 m^m (1-m)^{1-m})^N.
\end{equation}
For $M_0 =N/4$, this gives
\begin{equation}\label{zwoelf} 
\langle T_{neq}\rangle  \ge \sqrt{3}\; (3^{3/4}/2)^N 
  \approx \sqrt{3}\; \exp(0.131N)\ \ ; 
\end{equation} 
the exponent is close to the fit given in Eq. (\ref{elf}).
Together with the upper bound $\langle T_{neq}\rangle \le 2^N - 1$ one obtains
exponential scaling with $N$. We find that usually $T_{neq}$ is proportional
to the return time of the non-equilibrium state, with a proportionality
constant that depends on the details of the generator (for example, 
if a two-tap or four-tap register is chosen, see Fig. \ref{sh-inv}).
Fig. \ref{sh-ent} shows the time dependence of the Boltzmann 
entropy $S_B (x(t))$, averaged over initial states with $M= N/4$  
(non-equilibrium) and $M=N/2$ (equilibrium), respectively. 
The non-equilibrium entropy decays to its equilibrium value  
$S_B \simeq S_G$ in $N$ time steps, since every bit $x_i$ is
visited and ``randomized'' exactly every $N$th time step. 
Starting from equilibrium states, the entropy stays constant. 
Note that after the Poincar\'e return time $T_R$, $S_B$ has to  
return to its initial value. But the time to reach equilibrium 
is of the order of the microscopic time, whereas the  
return time $T_R$ increases exponentially with system size.  
\section{Stochastic Ehrenfest model} 
\label{StochEhr}
\begin{figure}
 \begin{center}
   \resizebox{0.95 \columnwidth}{!}{\includegraphics{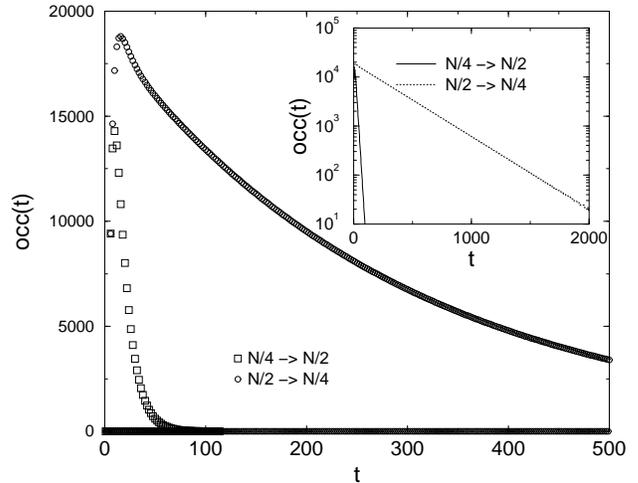}} 
 \end{center}
 \caption{Distribution of first-passage times in the 
stochastic Ehrenfest model, using the same normalization as
in Fig. \ref{sh-fpt}.}
\label{ef-fpt}
\end{figure}

The original Ehrenfest model describes a stochastic process:
$N$ balls are distributed among two urns. At every time step, a ball is
picked at random and moved to the respective other urn;
$M$ is the number of balls in one of the urns. This process can
be mapped to a random walk with drift \cite{Kac:RW}. 
$M$ is the position of the
walker, who at each time step performs a step $\Delta M = \pm 1$ with
probability
\begin{eqnarray}\label{dreizehn} 
P(M \rightarrow M+1) & = & (N-M)/M \nonumber \\ 
P(M \rightarrow M-1) & = & M/N \ \ . 
\end{eqnarray} 
Hence, there is a drift towards the center $M=N/2$. To our knowledge,
analytic calculations of first passage times exist only between
equilibrium states \cite{Lipowski:Absorption}, 
whereas we are interested in the decay to or
from non equilibrium states. We have calculated these times from
numerical iteration of the corresponding transition matrices using
absorbing states \cite{VanKampen}. The average times
for different system sizes are shown in Figs. 
\ref{sh-rel} and \ref{sh-inv} in
comparison with the corresponding deterministic model;
the distribution of times for $N=24$ is shown in Fig. \ref{ef-fpt}. 
Again, the decay
time $T_{eq}$ to equilibrium scales with $N$ (although with
more pronounced finite-size effects) while the time to reach a
non equilibrium state from an equilibrium one increases exponentially
with system size. Indeed the stochastic model has similar properties as
the deterministic one. 

\section{Hierarchical dynamics} 
\begin{figure}
 \begin{center}
   \resizebox{0.95 \columnwidth}{!}{\includegraphics{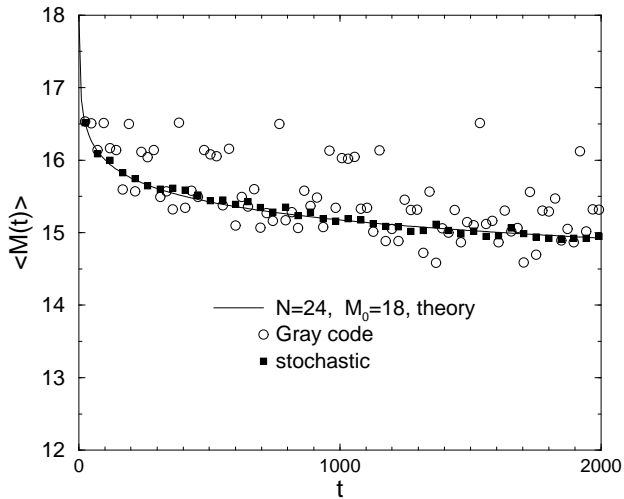}} 
 \end{center}
 \caption{Relaxation of $\langle M \rangle$ in the Gray code
   and corresponding stochastic model, compared to Eq. (\ref{vierzehn}). 
   Relaxation times are exponential in $N$.}
 \label{expmag}
\end{figure}

There are several deterministic algorithms which change only one bit 
per time step and
run through all of the $2^N$ configurations of the bit sequence $x$. One
of them is the Gray Code \cite{NumRec}: Find the largest 
power $2^j$ by which the
discrete time $t$ is divisible and change bit $x_{j+1}$. The quantity $M$
changes by $\pm 1$ only, and the system runs through all of the $2^N$
possible states. However, now the access to the different variables
$x_j$ is hierarchical; it takes $2^{j+1}$ steps before site $j$ is
visited again. This hierarchical structure of microscopic dynamics leads
to slow relaxation times of the macroscopic property $M$, as shown in
Fig. \ref{expmag}. The time to reach an equilibrium state from an 
initial non-equilibrium one scales exponentially with system size $N$.
 
\begin{figure}
 \begin{center}
   \resizebox{0.95 \columnwidth}{!}{\includegraphics{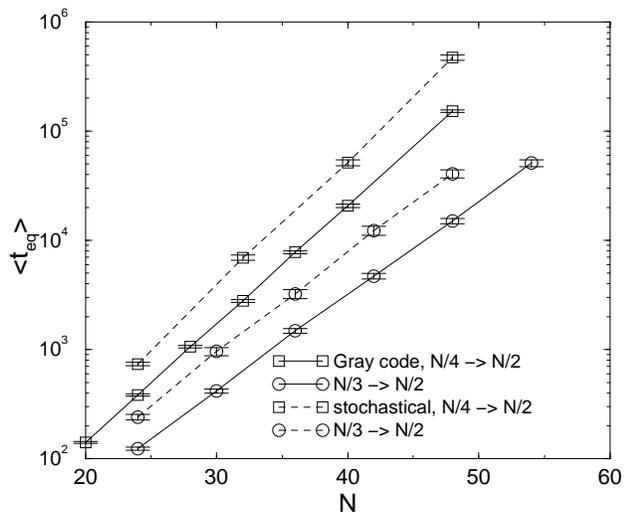}} 
 \end{center}
 \caption{In the Gray code and corresponding stochastical model, times for
transition to equilibrium are exponential in $N$, with identical
exponents but different prefactors.}
\label{exp-rel}
\end{figure}

There is a corresponding stochastic version of the Gray code: 
\begin{itemize} 
\item Start with $j=1;$ 
\item with probability $1/2$, increase $j$ by 1; else flip $x_j$,
increase $t$ by one and return to the first step;
\item if $j>N$, increase $t$ and return to the first step; else
repeat the second step. 
 \end{itemize} 
The ensemble average of $M$ decays as 
\begin{equation}\label{vierzehn} 
\langle M \rangle =  N/2 + 
   \sum\limits^N_{j=1} (M_0/N - 1/2) 
    \exp(- t/2^{j-1})  \ \ . 
\end{equation} 
The simulations of the stochastic model agree with Eq. (\ref{vierzehn});
however, in the deterministic Gray code, $\langle M \rangle$ shows structures
caused by ''almost returns'' after times that are multiples of high
powers of $2$ (see Fig. \ref{expmag}).  The first passage times of the 
stochastic model show a similar exponential scaling as the 
corresponding times of the deterministic model (see Fig. \ref{exp-rel}).

It is possible to have a quadratic dependence on $N$ of the
decay time of $\langle M(t) \rangle$ by the following stochastic 
algorithm:\\
Draw a uniform random number $r \in [0,1]$ and flip $x_j$ with the largest index $j < 1/r$, if $j\le N$.\\
The probability to flip $x_j$ is of order $1/j^2$. The average value of
$M(t)$ decays as
\begin{eqnarray} 
\langle M \rangle &\approx& N/2 + 
  (M_0/N - 1/2 )   [ (N+1) 
    \exp (-\frac{2t}{(N+1)^2} ) \nonumber \\ 
  & &   - \exp(-2t)
+ \sqrt{2 \pi t} (\erf(\frac{\sqrt{2t}}{N+1}) - \erf(\sqrt{2t})) ]   .
\end{eqnarray} 
The long term behaviour is determined by the longest decay constant,
which is $-2/(N+1)^2$. We have also calculated the first passage time to
equilibrium, and find that it scales with $N^2$.
\section{Summary} 
A simple model with time reversible deterministic dynamics is
investigated. The Poincar\'e return time is $2^N-1$, where $N$ is the size
of the system; each initial state returns to itself after $2^N-1$ time
steps. We define a macroscopic quantity $M(t)$ and study its time
dependence. When the system starts from an equilibrium state (state with
most probable value of $M$) it takes on average an exponentially large
time before it reaches a state far from equilibrium. On the other side,
when the system starts far from equilibrium it reaches an equilibrium
state after a period of the order of the microscopic time. The same is
true for the corresponding Boltzmann entropy. This behaviour explains
the time's arrow in macroscopic properties for a system with time
reversible dynamics.
 
We have calculated the full distributions of these first passage times
and compared them to the ones of the corresponding stochastic Ehrenfest
model. Surprisingly the distribution of times the system needs to
reach non-equilibrium for the first time reaches a
maximum at times of order $N$ and  monotonically decreases for larger
times. However, the integral of $T_{neq}$ over short 
times is small compared to the total number of equilibrium
states. Therefore, on average the time to leave equilibrium scales
exponentially with system size. If our computer needs $10^6$ steps per
second, Eq. (\ref{zwoelf}) gives the age of the universe for $N\approx400$.
Even for our simple model the return to non-equilibrium cannot be
observed in reasonable time for $N$ significantly larger than 70.     
 
All these results hold only for a model with identical microscopic times
for all of the local variables. We have also studied models with
hierarchical or power law access times. In these cases the relaxation to
equilibrium is slowed down.  The time to reach equilibrium scales
exponentially or with a power of system size, depending on the
distribution of microscopic times.

\section{Acknowledgement}
All authors are grateful for financial support by the German-Israeli 
Foundation. This work also benefitted from a conference at the
Max-Planck Institut f\"{u}r Physik komplexer Systeme, Dresden.

\end{multicols}


\begin{thebibliography}{10}

\bibitem{Lebowitz:Macroscopic}
J. Lebowitz, Physica A {\bf 1994},  1  (1993).

\bibitem{Lebowitz:Entropy}
J. Lebowitz, Phys. Today {\bf 46},  32  (1993).

\bibitem{Lebowitz:More}
J. Lebowitz,  in {\em More Things in Heaven and Earth}, edited by B. Bederson
  (Springer Verlag, New York, 1999), p.\ 581.

\bibitem{Stauffer:Q2R}
D. Stauffer, Comp. Phys. Commun. {\bf 127},  113  (2000).

\bibitem{Ambegaokar:Entropy}
V. Ambegaokar and A. Clerk, Am. J. Phys. {\bf 67},  1068  (1999).

\bibitem{Lebowitz:Microscopic}
J. Lebowitz, Physica A {\bf 263},  516  (1999).

\bibitem{Prigogine:Laws}
I. Prigogine, Physica A {\bf 263},  528  (1999).

\bibitem{Ehrenfest:Model}
P. and T. Ehrenfest, Phys. Zeit {\bf 27},  311  (1908).

\bibitem{Kac:Probability}
M. Kac, {\em Probability and Related Topics in Physical Sciences} (American
  Mathematical Soc., Providence, R.I., 1959).

\bibitem{Lebowitz:Lorentz}
J. Lebowitz and H. Spohn, J. Stat. Phys. {\bf 19},  633  (1978).

\bibitem{Lebowitz:Lorentz2}
J. Lebowitz and H. Spohn, J. Stat. Phys. {\bf 29},  39  (1982).

\bibitem{Levesque:Dynamics}
D. Levesque and L. Verlet, J. Stat. Phys. {\bf 73},  512  (1993).

\bibitem{Knuth:Art2}
D. Knuth, {\em Seminumerical Algorithms}, Vol.~2 of {\em The Art of Computer
  Programming}, 2nd  ed. (Addison-Wesley, Redwood City, 1981).

\bibitem{NumRec}
W. Press, S. Teukolsky, W. Vetterling, and B. Flannery, {\em Numerical Recipes
  in C}, 2nd  ed. (Cambridge University Press, Cambridge, 1992).

\bibitem{Schroeder:Zahlentheorie}
M. Schroeder, {\em Number Theory in Science and Communication}
  (Springer-Verlag, Berlin, 1984).

\bibitem{Ziff:Four-tap}
R.~M. Ziff, Computers in Physics {\bf 12},  385  (1998).

\bibitem{Kac:RW}
M. Kac, Amer. Math. Monthly {\bf 54},  369  (1947).

\bibitem{Lipowski:Absorption}
A. Lipowski, J. Phys. A {\bf 30},  L91  (1997).

\bibitem{VanKampen}
N. van Kampen, {\em Stochastic Processes in Physics and Chemistry} (Elsevier
  Science B.V., Amsterdem, 1992).

\end{thebibliography}


\end{document}